\documentstyle[12pt]{article}
\evensidemargin\oddsidemargin
\begin{document}
\count0 = 1
\begin{titlepage}
\vspace {20mm}
\begin{center}
\ QUANTUM REFERENCE FRAMES\\
\smallskip
 AND RELATIVISTIC TIME OPERATOR  \\
\vspace{6mm}

\bf{S.N. MAYBUROV}
\vspace{6mm}

\small{Lebedev Institute of Physics}\\
\small{Leninsky pr. 53, Moscow Russia, 117924}\\

\vspace{15mm}
\end {center}
\vspace{3mm}
\begin {abstract}
Aharonov-Kaufherr model of quantum space-time which accounts 
  Reference Frames (RF) quantum effects is considered in Relativistic 
  Quantum Mechanics framework. For RF  connected with some 
macroscopic  object  its free quantum motion -
wave packet smearing results in 
additional uncertainty of test particle coordinate.
Due to the same effects the use of
Galilean or Lorentz transformations for this RFs becomes incorrect and
 the special quantum space-time transformations are introduced.
 In particular for any RF the proper time becomes the operator 
in other RF. This time operator calculated solving 
relativistic Heisenberg equations for some quantum clocks models.
 Generalized Klein- Gordon equation proposed
 which depends on both the particle and  RF masses.  
\end {abstract}
\vspace{24mm}
\small {Submitted to Phys. Lett. A }\\
\vspace {28mm}
\vspace {10mm}
\small {Talk given at 'Uncertain Reality' simposium
 , New Delhi , January 1998}\\
\vspace{28mm}
\small {  * E-mail  Mayburov@sgi.lpi.msk.su}
\end{titlepage}
\begin{sloppypar}
\section{Introduction}

Some years ago Aharonov  and Kaufherr have shown that in nonrelativistic
Quantum Mechanics (QM) the correct definition of physical reference 
frame (RF) must differ from commonly accepted one, which in fact was
transferred copiously from Classical Physics $\cite{Aha}$. The main reason  
 is that to perform exact quantum description
 one should account the quantum properties 
not only of studied object, but also RF, despite the possible practical
 smallness.   
 The most simple  of this RF properties is the existence of Schroedinger
 wave packet of free macroscopic object with which RF is usually
associated $\cite {Schiff}$. If this is the case it inevitably introduces
additional uncertainty in the measurement of object space coordinate.
Furthermore  this effect account results in the coordinate transformations
between two such RFs called quantum RFs,
  principally different from the Galilean ones $\cite{Aha,Tol}$.

In their work Aharonov and Kaufherr formulated Quantum Equivalence 
principle - all the laws of Physics are invariant under transformations
between both classic and quantum RFs. In their paper
  its applicability for nonrelativistic QM was proved.
  
 The importance of RF quantum properties account was shown already in 
Quantum Gravity and Cosmology studies  $\cite {Rov,Unr}$.
Further studies of quantum RF effects  can help also
 to understand some features of quantum space-time 
at small distances  $\cite{Dop}$. 
 The aim of our study is
the development of  relativistically covariant  quantum RF description ;
our first results were reported in \cite{May2}.
 It will be shown that the transformations of the test particle state vector
 between two quantum RF obeys to relativistic invariance
principles, but due to dependence on RF state vector
differs from Poincare Group
transformations. The time ascribed to such RF becomes the operator 
, corresponding to proper time of Classical Relativity.
 As will be shown this operator introduces
 the quantum fluctuations in
the classical Lorentz time boost in moving RF time measurements.
Our paper is organized as follows : in the rest of this chapter
 our model of quantum RF will be formulated and its compatibility with
 Quantum Measurement Theory discussed.
In a chapter 2 the new canonical formalism of quantum RF states and their
transformations  described, which is quite  simple and more suitable to our
purposes.
  The relativistic equations for quantum RF
and the resulting quantum space-time transformations are regarded in 
chapter 3 . In a final chapter
the obtained results and their interpretation are discussed.
 
 In QM framework the system regarded as RF presumably should be able
 to measure 
the observables of studied quantum states and due to it
 to  include measuring devices - detectors.
 As the realistic example of such RF we can regard the photoemulsion plate or
the diamond crystal which can measure microparticle position relative to its
c.m. and simultaneously record it.
At first sight it seems that  due to it quantum RF problem must use  
as its basis some model of the state vector collapse.
 Yet despite the multiple proposals up to now well established
theory of collapse doesn't exist  $\cite{May,Desp}$.
  Alternatively we'll show that our problem premises doesn't connected
 directly with the state vector collapse mechanism and
 and in place of it the two simple assumptions
about  RF and detector states properties can be used.
  The first one is that  RF consists of 
 finite number of atoms (usually rigidly connected)  and have the finite
 mass.
 Our second assumption needs some preliminary comments.
 It's well known that the solution of Schroedinger equation for
  any free quantum system  consisting of $N$
constituents can be presented as :
\begin {equation}
  \Psi(\vec{r}_1,...,\vec{r}_n,t)=
\sum c_l\Phi^c_l(\vec{R}_c,t)*\phi_l(\vec{r}_{i,j},t) \label {A1}
\end {equation}
where  center of mass coordinate $\vec{R}_c=\sum m_i*\vec{r}_i/M$.  
 $\vec{r}_{i,j}=\vec{r}_i-\vec{r}_j$  are  the relative or 'internal'
 coordinates of constituents
 $\cite{Schiff}$. Here $\Phi^c_l$ describes the c.m. motion of the system.
 It means that the evolution of the
 system  is separated into the external
evolution  of  pointlike particle M and the internal evolution
 completely defined by $\phi_l(\vec{r}_{ij},t)$
 So the internal evolution is
 independent of whether the  system is localized
 in the  macroscopic 'absolute' reference frame (ARF)
 or not. Relativistic QM and Field Theory  evidences that the
 factorization of c.m.
 and relative motion holds true even for nonpotential forces and 
 variable $N$ in the secondarily quantized systems $\cite{Schw}$.
 Moreover this factorization expected to be correct for nonrelativistic
 systems 
 where binding energy is much less then its mass $M$, which is
 characteristic for the real detectors and clocks.
 Consequently it's reasonable
to extend this result on the detector states despite we don't know
their exact structure. We'll use it quite restrictively and 
  assume that
   the factorization of the c.m. motion holds for RF
  only in the time interval $T$
 from RF preparation moment , until the act of measurement starts
,i.e. when the measured particle collides with it. Formally our second 
  assumption about RF properties is    
that  during period $T$ its state is described by  wave function
generalizing  (1) :
\begin {equation} 
 \Psi(\vec{R}_c,u_i,t)=\sum c_l\Phi^c_l(\vec{R}_c,t)
*\varphi_l(u_i,t) \label {A2}
\end {equation}
 where $u_i$ denote all internal detector degrees of
freedom, and this state
  evolves during $T$ according to Schroedinger equation
(or some field equation). Its possible violation at later time when
the particle state collapse occurs is unimportant for our model.
 Due to it we'll assume 
always  in our model that all measurements  are performed on
the  quantum ensemble of observers $F^1$. It means that each event is resulted 
 from the interaction between the 'fresh' RF and particle
 ,prepared both in the specified quantum
 states ,alike the particle alone in the standard experiment.
 To simplify our calculations normally we'll  take below all $c_l=0$
 except $c_1$ which wouldn't influence our final results.

The common opinion is that to observe experimentally 
measurable smearing of macroscopic object
demands too large time , but for some mesascopic experiments
it can be reasonably small to be tested in the laboratory conditions
$\cite{May2}$.  
 We don't consider in our study the influence of RF recoil
 effects on the measurements results which can be made arbitrarily
small $\cite {Aha}$.  

\section{Quantum Coordinates Transformations }
To illustrate the meaning of Quantum RF
 consider gedankenexperiment in two dimensions $x,y$, where in ARF
the wave packet of RF $F^1$  
  described by  $\psi_1(x)\xi_1(y)$ at some time moment $T$. 
 The test particle $n$
  with mass $m_2$ belongs to   narrow beam which average velocity
is  orthogonal to $x$ axe
 and  its wave function at $T$ is $\psi_n(x)\xi_n(y)$. Before they start
to interact this system wave function is the product of $F^1$ and $n$
packets.
We want to find $n$ wave function for the observer in
 $F^1$ rest frame. In general it can be done by means of the canonical 
transformations described  below, 
but in the  simplest case when $n$ beam is  localized  
and $\psi_n(x)$ can be approximated
 by  delta-function $\delta(x-x_b)$ this wave function in $F^1$ as easy to find
$\psi'_n(x_n)=\psi_1(x_n-x_b)$. It shows that if $F^1$ wave packet
have average width $\sigma_x$ then from the 'point of view'
of observer in $F^1$ each object localized in ARF acquires wave packet of the
same width $\sigma_x$ and any measurement both in $F^1$ and ARF
, as will be shown below confirms this conclusion.
 
Now we'll regard
the  nonrelativistic formalism 
alternative to  Quantum Potentials one used in $\cite{Aha}$.
 Consider the system $S_N$ of $N$ objects $B^i$  which
 include $N_g$ pointlike 'particles' $G^i$ and $N_f$ frames $F^i$,
which in principle can have also some internal
 degrees of freedom described by (1).
For the start we'll take that particles and RF coordinates 
$\vec{r}_i$ are given
 in absolute (classical) ARF having very large
mass $m_A$.
We should find two transformation operators - from ARF to quantum RF
,and between two quantum RF, but it'll be shown that in  general 
approach  they coincide .
We'll use Jacoby canonical
coordinates  $\vec{q}_j^l$, which for $F^l$ rest frame are equal :
\begin{eqnarray}
 \vec{q}^l_{i}=\frac{\sum\limits^N_{j=i+1}m^l_j\vec{r}^l_{j}}{M^{l,i+1}_N}
-\vec{r}^l_{i} , j<N ;
 \quad \vec{q}^l_N=\vec{R}_{cm} \quad \label{B1}
\end{eqnarray}
Here $\vec{r}^1_j=\vec{r}_j , m^1_j=m_j$ , and  for $l>1$ : 
\begin {equation}
 \vec{r}^l_1=\vec{r}_l ;\quad
\vec{r}^l_j=\vec{r}_{j-1} , 1< j\leq l ;\quad 
\vec{r}^l_j=\vec{r}_j , j>l \label{B1A}
\end {equation}
The same relations connect $m^l_j$ and $m_j$,
  $M^{l,i}_n=\sum\limits^{n}_{j=i}m^l_j$ (if upper indexes $i,l$ are omitted,
then $i,l=1$).
 Conjugated to $\vec{q}^l_i$ canonical momentums 
can be easily found, for example  :
\begin {equation} 
\vec{\pi}^1_i=
\mu_i(\frac{\vec{p}^s_{i+1}}{M^{i+1}_N}-\frac{\vec{p}_{i}}{m_{i}}) ,\quad
\vec{\pi}^1_N=\vec{p}^s_1 \label{B2}
\end {equation}
where $\vec{p}^s_i=\sum\limits^{N}_{j=i}\vec{p}_j$
,and reduced mass
 $ \mu^{-1}_i=(M^{i+1}_N)^{-1}+m_{i}^{-1} $ .

The relative coordinates  $\vec{r}_j-\vec{r}_1$ can be represented
 as the linear sum of several coordinates
 $\vec{q}^1_i$ ; they don't
constitute canonical set due to the quantum motion of $F^1$.

 We consider first
 the transformation between two quantum RF and start from
the simplest case $N_f=2,N_g=0$. This is just the space reflection
of  $F^1$ coordinate $\vec{q}^2_1=-\vec{q}^1_1$ performed
 by the parity operator $\hat{P}_1$. 
 The next case $N_f=2,N_g=1$ is 
 $\vec{q}^1$ coordinates bilinear transformation exchanging $\vec{r}_2,
\vec{r}_1$ :
\begin {equation}
  \vec{q}^2_{1,2}=\hat{U}_{2,1}\vec{q}^1\hat{U}^{+}_{2,1}
=a_{1,2}\vec{q}^1_1+b_{1,2}\vec{q}^1_2 \label {B3}
\end {equation}
Corresponding unitary operator can be decomposed as 
$\hat{U}=\hat{C_2}\hat{R}\hat{C_1}$ ,where $\hat{C_{1,2}}$
are the dilatation  operators,
 which action changes the coordinate scale. For example $\hat{C}_1$
results in $\vec{q}^1_i=c^i_{1}\vec{q}^1_i$, where
 $c^i_{1}$ proportional to $\mu_i$.

 $\hat{R}$ is the rotation on $\vec{q}^1_{1,2}$ intermediate
coordinates hypersurface on the angle :
\begin {equation}
\beta=-\arccos[\frac{m_2m_1}{(m_3+m_2)(m_1+m_3)}]^\frac{1}{2}
\end {equation} \label {B3A}
 For the general case $N>3$ it's possible nevertheless
to decompose the transformation from $F^j$ to $F^k$ 
as the product of analogous bilinear operators. Really if to denote
as $\hat{S}_{i+1,i}$ the operator exchanging $F^i,F^{i+1}$  in $\vec{q}^1$ set,
as follows from (\ref{B1}) it changes in fact only
$\vec{q}^1_i,\vec{q}^1_{i+1}$ pair.  $\hat{U}_{2,1}=\hat{S}_{2,1}$
and all $\hat{S}_{j,j-1}$ have the analogous form  changing only parameters
$\beta,c^i_k$.   
Then the transformation operator
 from $F^1$ to $F^k$ is :
\begin {equation}
   \hat{U}_{k,1}=\hat{S}_{2,1}\hat{S}_{3,2}...\hat{S}_{k,k-1} \label {B4}
\end {equation}
It follows immediately that the transformation from $F^j$ to $F^k$ is
 $\hat{U}_{j,k}=\hat{U}_{k,1}\hat{U}^{-1}_{j,1}$.

To find the transformation operator from the classical ARF
 to $F^1$ we'll regard ARF as the quantum  object $B^{N+1}$
with infinite $m_{N+1}$ belonging to extended system
 $S_{N+1}$. 
ARF 'classical' Jacoby set is $\vec{q}^A_i=\vec{r}_{i}-\vec{r}_A$ , but acting
by parity operators we'll transform it to $\vec{q}^A_i=-\vec{q}^A_i$.
 Then it's easy
to see that for $S_{N+1}$ $\vec{q}^1_i=\vec{q}^A_i$ as follows from
 (\ref{B1}). 
 Note that formally we can regard also each
particle  $G^j$ as RF and perform for them the transformations
 $\hat{S}_{j,j-1}$ described above. 
Then omitting simple calculations we obtain that operator
performing transformations from ARF to $F^1$ 
 is equal to $\hat{U}_{A,1}=\hat{U}_{N+1,1}$ for infinite $m_{N+1}$.
In this case  new $\vec{q}^l$ set  for $S_{N+1}$ 
 can be rewritten as the function of $\vec{r}^l_i,
\vec{r}^l_{N+1}=\vec{r}^l_A$  of (\ref{B1}) to which formally must be added
 $\vec{q}^{l}_{N+1}=\vec{r}_A$.

 The free Hamiltonian of the system objects motion in ARF is   :
\begin {equation} 
\hat{H}=\hat{H}_s+\hat{H}_c=\frac{(\vec{\pi}^{1}_N)^2}{2M_N}+
\sum\limits_{j=1}^{N-1}\frac{(\vec{\pi}^{1}_i)^2}{2\mu_i} \label {B5}
\end {equation}
Hamiltonian of $S_N$ in $F^1$ should depend on relative $B^i$ momentums
only , so we can regard $\hat{H}_c$ as the real candidate for its role.
It results into modified Schroedinger equation, in which objects evolution
depends in fact on observer mass $m_1$.
Yet relativistic analysis given below introduces corrections to
 $\hat{H}_c$, which normally are small but essential for 
the interpretation.

 In general this quantum transformations in 2 or 3 dimensions 
should also take into account the possible
 rotation of quantum RF relative to ARF, which introduce 
additional angular uncertainty into objects coordinates.
  Thus after
 performing coordinate transformation $\hat{U}_{A,1}$ from ARF
to $F^1$ c.m.  we must rotate all the 
objects (including ARF) around it on the uncertain  polar and
azimuthal angles $,\phi_1,\theta_1$
,so the complete transformation is
 $\hat{U}^T_{A,1}=\hat{U}^R_{A,1}\hat{U}_{A,1}$.
Such rotation transformation operator commutes with $\hat{H}_c$ and due to it
can't change the evolution of the transformed states $\cite{May2}$.

Now we'll discuss the measurements in quantum RF and for this purpose 
return to the gedankenexperiment regarded above. 
Let's assume in addition that $F^1$ includes
detector $D_0$ which can measure the distance between $n$
and $F^1$ c.m. $\Delta=x_n-x_1$.
 $F^1$ and ARF observers will treat the same event unambiguously
as  $n$ measurement by $D_0$  (or its flight through $D_0$). In
 $F^1$  $n$ state vector collapse
reveals itself by the detection  process in $D_o$
 initiated by $n$ absorption. For ARF
the collapse results from the $n$ nonobservation  in a due time
in ARF detectors  - so called negative result experiment.
 This measurement means not only the reduction of $\psi'_n$ in $F^1$
to some $|\Delta_j\rangle$ eigenstate, but also the reduction of $\psi_1(x)$
in ARF to $\delta(x-x_b)$ in our example.
Now let's consider for arbitrary $\psi_n$ the reduction in ARF for
$\Delta$ measurement of the state 
$\psi_s=\psi_n(x_n)\psi_1(x_1)\varphi(u_i)$ corresponding to the
 density matrix $\hat{\rho}_{in}$, where $\varphi$ is
$F^1$ internal (detector) state. If the measurement of  $|\Delta_j\rangle$
 eigenstate results in
  $\varphi_j$ detector state, described correspondingly 
by the projection operators $\hat{P}_{\Delta j},\hat{P}_{\varphi j}$
, then the density matrix after the collapse is given by  \cite{Hol} :
\begin {equation} 
\hat{\rho}^1_f=\sum_j \hat{P}_{\Delta j}\hat{P}_{\varphi j}\hat{\rho}^1_{in}
\hat{P}_{\Delta j}\hat{P}_{\varphi j} \label {B11}
\end {equation} 
In $S_2$ Jacoby coordinates $\Delta=q^1_{1x} , x_{cm}=q^1_{2x}$. Then
from the relation $|\Delta\rangle |x_{cm}\rangle=|x_1\rangle\ |x_\rangle$
, it follows that $\hat{\rho}^1_f$ is a mixture of 
the states nonlocalized in ARF
of the form $\psi_n(x)\psi_1(x-\Delta_j)$ for the given $\Delta_j$ value.
 Their
effective width is of the order $\sigma_n$, if $\sigma_n \ll \sigma_1$.
 This $F^1$ state reduction
takes place for any $\psi_n$, but without $F^1$ localization
in ARF.
So in general $n$ coordinate $\Delta$ measurement in $F^1$ transforms initial 
$F^1$ state into another nonlocalized state. This results demonstrates
that nonlocalized RF conserves this delocalization after the measurements
of the other nonlocalized states.
 The Decoherence model 
calculations of the particle coordinate measurements  by $F^1$
detector support this conclusion \cite{May2}.

\section  {Relativistic Equations}

Now we'll consider possible generalization of
Quantum Equivalence principle for relativistic QM.
The relativistic covariant formalism will be studied here 
with the model of  relativistic wave packets
of macroscopic objects  regarded as quantum RF $\cite{Blo}$.  
 We'll take that all RF 
   constituents spins and orbital momentums are  compensated 
so that its total orbital momentum is zero.

In nonrelativistic mechanics time $t$ is universal and 
 is independent of observer, while in relativistic
case  each observer in principle has its own proper time $\tau$.
We don't know yet the origin of the physical time , but phenomenologically 
 we can associate it with the clock hands motion or some other relative
motion of the system parts.
Meanwhile  it will be
 shown that like in the case of the position measurement this internal
 processes can be disentangled from the system c.m. motion.
 Then RF $F^2$ wave packet evolution can be described by the
 relativistic equation 
  for their c.m. motion relative to  other RF and $F^2$ internal degrees of
freedom evolution which define its clocks motion 
and consequently its proper time $\tau_2$ are factorized from it.
 
We'll start the proper time study with the simple models
of quantum RFs with clocks, yet we expect its main results to be true
also for the more sophisticated models.  
Consider the evolution of some system $F^2$ where the  internal interactions
are nonrelativistic , which as was discussed  in chap.1 is a reasonable
 approximation for the measuring devices or clocks. More precisely
we consider the complex system which Hamiltonian in ARF is analogous to
Klein-Gordon square root (KGR) Hamiltonian of pointlike particle \cite{Schw} :
$$
   {H}_T=[(m'_2+H_c)^2+\vec{p}_2^2]^{\frac{1}{2}}
$$
, where $H_c=V+T$ is clocks Hamiltonian , $m'_2$ is $F^2$ constituents rest
 mass.
We'll use below the parameter $\alpha_I=\frac{\bar{H}_c}{m'_2}$ , describing
the relative strength of the internal $F^2$ forces , which for the
realistic clocks can be as small as $10^{-10}$. In $F^2$ c.m. it can be
defined as  
 $\alpha_I=m'^{-1}_2\langle\varphi_2|\hat{H}_c|\varphi_2\rangle$
 where $\varphi_2$ is $F^2$ internal state of ($\ref{A2}$) , yet our general
 formalism doesn't demand factorizable $\varphi_2$ existence at all.  

As the illustrative example
we'll consider the model of quantum clocks - rotator $C_q$ proposed by Peres
\cite{Per} with the Hamiltonian 
 $\hat{H}_c=-2\pi \omega i \frac{\partial}{\partial\theta}$
,where $\theta$ is the rotator's  polar angle ($-\pi<\theta<\pi$).
Preparing the special $C_q$ initial state at $t=0$, which is analog
of Gaussian wave packet :
$$
 |v_0\rangle=\sum_{m}\frac{e^{im\theta}}{N^\frac{1}{2}}
$$
where $m=-J_z,...,J_z,~N=2J_z+1$ one obtains the close resemblance
 of the classical clocks evolution.
Resulting $C_q$ state $\varphi_c(\theta-2\pi\omega t)$
 for large $N$ has the sharp peak at
$\bar{\theta}=2\pi\omega t$ with the uncertainty
 $\Delta_{\theta}=\pm \frac{\pi}{N}$ and can be visualized as the constant 
hand motion on the clocks circle. As was shown in \cite{Hol} 
the observable $\hat{\tau}$ performs the nonshifted measurement
 of time parameter $t$ if at any $t$
 $\bar{\tau}=t$ and its dispersion $D(\tau)$ is finite.
Then the operator 
$\hat{\theta}_{R}=\frac{\theta}{2\pi\omega}$ describes
 nonshifted $t$ measurement in the interval $0\leq t < T$, where 
$T=\omega^{-1}$
where its dispersion $D(\theta_R)=\frac{T^2}{4N^2}$
, (for $t>T$
it describes $t'=mod(t,T)$). 
 Suppose that both  $F^2$ and ARF carry the clocks $C_q, C'_q$ 
performing the proper time measurements for the observables 
$\hat{\tau}_2^p=\hat{\theta}_R , \hat{\tau}_0^p=\theta'_R$ in their rest 
frames.
Yet $C_q$ angle $\theta$ in principle can be measured also in ARF
 so that
its proper time $\hat{\tau_2}$ is the observable in ARF
   proportional to  $C_q$ angle $\theta$ and differ from $\hat{\tau}_2^p$.
For the simplicity we'll suppose below that
$D(\theta_R)\gg D(\theta'_R)$ and $T'=T$.

This relativistic time operator $\hat{\tau}_2$ can be
 found solving  Heisenberg  equation  for the $C_q$ 
relativistic Hamiltonian $\hat{H}_T$. 
 In particular to obtain $C_q$ angle operator
$\theta(\tau_0)$ for Hamiltonian $H_T$ in ARF 
  the commutation
relations for operator functions
$ [Q,F(P)]=-i F'(P) $  can be used \cite {Hol}. 
 After the simple algebra one obtains  the  evolution equation
for $\theta(\tau_0)$ :
\begin {eqnarray}
  \dot{\theta}=-i[\theta,H_T]=
\frac{2\pi\omega m_2}{(m_2^2+\vec{p}_2^2)^\frac{1}{2}}
=2\pi\omega B_2 
\label {C0H}
\end {eqnarray}
where $m_2=m'_2+H_c$ is $F^2$ mass operator.
The solution of this equation is : \\
 $\theta(\tau_0)=2\pi\omega B_2\tau_0+\theta(0)$
where $B_2$ can be called time boost operator.
 From the relation
$\hat{\tau_2}=\frac{\theta}{2\pi\omega}$  
 it follows that the proper time operator is equal to : 
\begin {equation}
   \hat{\tau_2}=B_2\tau_0+\frac{\theta(0)}{2\pi\omega} \label {C2W}
\end {equation}
where constant operator $\theta(0)$ can be defined
from $C_q$ nonrelativistic Hamiltonian $H_c$.
 Despite $\bar{\theta}(0)=0$ it produces the additional
quantum fluctuations. Due to the ciclical $C_q$ evolution
  equation (\ref{C2W}) is formally fulfilled only for $\tau_0<T$  
, but below it will be shown that it has fundamental origin
independent of the paricular clocks model and holds for any $\tau_0>0$.

Then $\hat{\tau}_2$ expectation value and dispersion are : 
\begin {eqnarray}
\bar{\tau}_2=\bar{B}_2(\vec{p}_2)\tau_0  \quad \label {C2X} \\
D(\tau_2)=D(B_2)\tau_0^2+\bar{G}_2\tau_0+D_0 \nonumber
\end {eqnarray}
where  $D_0=D(\theta_R)$ and
$D(B_2)=\bar{B}^2_2-(\bar{B}_2)^2$. 
Operator $G_2$ in the second $\alpha_I$ order is equal to :
$$
G_2=\frac{2B_2\theta(0)}{2\pi\omega}
+\frac{\vec{p}_2^2(H_c\theta(0)+\theta(0)H_c)}
{4\pi^2\omega^2(m'^2_2+\vec{p}_2^2)^\frac{3}{2}}
$$
 $G_2$ is connected in fact with the interference between 
$B_2$ and $\theta(0)$ operators. 
It seems to be some analogy between this term and nonexponential and
momentum dependent corrections to the relativistic decay amplitudes \cite{Kha}. 

To make  the interpretation of the time operator $\hat{\tau}_2$
more clear let's consider first $C_q$ evolution in its rest frame.
Then from Heisenberg equation for $\theta , H_c$ it follows 
$\hat{\tau}^p_2=I\tau_2+\theta_R(0)$ , where $I$ is unit operator.
$\hat{\tau}_2^p$ dispersion is equal to $D_0$ which means that its
real difference from $I\tau_2$ can be done very small. $I\tau_2$ 
represents the parameter $\tau_2$  and $\hat{\tau}_2^p$ in fact is
pseudotime observable which approximates $\tau_2$ with the arbitrary 
 accuracy. In distinction $\hat{\tau}_2$ dispersion grows unrestrictedly
,which means that it can't aprroximates any time parameter.  
Consider now
 the evolution equation for $F^2$ state 
 in the first order of $\alpha_I$ , where it's possible to
 factorize total $F^2$ Hamiltonian $\hat{H}_T$ :
\begin {equation}
-i\frac{d\Psi_2}{d\tau_0}=(m_2^2+\vec{p}_2^2)^{\frac{1}{2}}\Psi_2 \simeq
[\frac{m'_2\hat{H}_c}{(m'^2_{2}+\vec{p}_2^2)^{\frac{1}{2}}}+
(m'^2_2+\vec{p}_2^2)^\frac{1}{2}]\Psi_2                \label {C01}
\end {equation}
For the simplicity we'll choose the special initial $F^2$ state
 $\Psi_2(0)=\Phi_2(\vec{p}_2)\varphi_2$ , where
$\varphi_2=|v_0\rangle$
 , such that
$C_q$ orbital momentum $J_z$ and $F^2$ momentum are parallel.
So $\Phi_2=\sum c_l|\vec{p}_{2l}\rangle$, and
 $\vec{p}_{2l}= (0, 0, p_{zl})$. This state describes $F^2$ clocks
sinchronized with ARF clocks at $\tau_0=0$. 
 It  evolves into
\begin {equation}
\Psi_2(\tau_0)=\sum c_l\varphi_{2l}(u_i,\tau_0)
|\vec{p}_{2l}\rangle e^{-iE(\vec{p}_{2l})\tau_0} \label {C0A}
\end {equation}
,where $\varphi_{2l}(u_i,0)=\varphi_2(u_i,0)$,
$E(\vec{p})=(m'^2_{2}+\vec{p}^2)^\frac{1}{2}$.
In this case  $\varphi_{2l}$ evolution in ARF is described by the
boosted Schroedinger equation :
\begin {equation}
-i\frac{d\varphi_{2l}}{d\tau_0}
=\frac{m'_2\hat{H}_c}{(m'^2_2+\vec{p}_2^2)^\frac{1}{2}}\varphi_{2l}=
\hat{B}_0(\vec{p}_{2l})\hat{H}_c\varphi_{2l}      \label {C02}
\end {equation}
This  equation describes the time dilatation in ARF in comparison with
$F^2$ c.m. for any processes in which $F^2$ constituents interact.
Due to this factorization it
 eq. (\ref{C02}) is easily solved for $F^2$ clocks :
\begin {equation}
\Psi_2(\tau_0)=\sum_l c_l \varphi_c(\theta-2\pi\omega B_{l}\tau_0)
~|\vec{p}_{2l}\rangle e^{-iE(\vec{p}_{2l})\tau_0} \label {C0B}
\end {equation}
where $B_{l}=B_0(\vec{p}_{2l})$. It shows that at any $\tau_0>0$  
$\Psi_2$ is the entangled superposition of the states which $F^2$ clocks
acquires at the consequent $\tau_2$ moments.
 It means for example that in ARF $F^2$ clocks
 can show 3,4 and 5 o'clocks simultaneously 
which can be checked by its hand angle measurement.
Note that $\hat{B}_2$ approximates the classical Lorentz factor
inverse value - $B_2=\gamma^{-1}(\vec{v})$, if the $\Phi(\vec{p}_2)$
packet width $\sigma_p\rightarrow0$. In this case one obtains
$\hat{\tau}_2=\gamma^{-1}I\tau_0+\theta_R(0)$ which gives just
classical time boost in moving RF for $\tau_2$ value.

Obtained results suppose that the proper time of any quantum RF
($F^2$) being the parameter in it simultaneously
 will be the operator from the 'point of view' of  other quantum RF.
This operator measurement shows how much time passed in $F^2$ 
in this particular event and can give quite different value
for another event.    
It means that the time moments in different RFs corresponds only
statistically with the dispersion of $\tau_2$ point in ARF 
given by (\ref {C2X}). It differs from Classical Relativity
where one -  to - one correspondence between $\tau_2 , \tau_0$ 
time moments always exists.

In general case $C_q$ state is quite complicated due to  Lorentz
transformation of the large orbital momentum components of $|v_0\rangle$.
But as follows from ($\ref{C2W}$)   
$\hat{\tau}_2$ expectation value and the dispersion leading term
are independent on it and
this state can influence only $D_0$ and $\bar{G}_2$ enlarging  so
 the clocks dispersion.

The more appropriate $C_x$ model of the time measurement
considers  the  free particle $m$ motion
 for the time
observable $\hat{t}=\frac{m x}{\bar{p_x}}$ proportional to 
the particle's path length  . For the Gaussian packet
$\varphi_x=A exp[-(\bar{p}_x a_x-p_x a_x)^2]$
the operator $\hat{t}$ in $F^2$ c.m.
 performs the nonshifted $t$ measurement with
 the finite dispersion for $0<t<\infty$ \cite {Hol}.
In the relativistic case we'll start with the Hamiltonian of
two objects $a,b$ relative motion ( see eq. ($\ref{C2}$) below )
in their c.m.s. :
$$
  \hat{H}_s=(m_a^2+\vec{q}_{ab}^2)^{\frac{1}{2}}
+(m_b^2+\vec{q}_{ab}^2)^{\frac{1}{2}}
$$
 where $\vec{q}_{ab}$ is $m_b$ relative invariant momentum \cite{Coe}. 
If $|\bar{q}_{ab}|$ is small we can choose as $p_x$  $\vec{q}_{ab}$ 
projection along any suitable direction and 
$x=i\frac{\partial}{\partial p_x}$. Then
 $\hat{H}_T$ mass operator
 $m_2=m_a+m_b+\frac{p_x^2}{2\mu_{ab}}+E_k(p_y,p_z)$ ,where $E_k$
can be neglected in the calculations. So the time in this model
can be defined measuring the distance between $F^2=a$ and some particle
$b$ emitted by $F^2$.

 Analogous to (\ref{C0H})  evolution equation results into
  the proper time operator :
\begin {equation}
\hat{\tau}_2=\frac{p_x B_2(\vec{p}_2)}{\bar{p}_x}\tau_0
+\frac{\mu_{ab} x(0)}{\bar{p}_x} \label {C2XX}
\end {equation}
Its expectation value and dispersion are given by (\ref {C2X}), but
$G_2$ and dispersion parameters are different :
\begin {eqnarray}
G_2= x(0)\frac{p_xm_2}{(m_2^2+\vec{p}_2^2)^\frac{1}{2}}
+\frac{p_xm_2}{(m_2^2+\vec{p}_2^2)^\frac{1}{2}}x(0)
 \nonumber \\  
D(B_2)=\frac{\bar{p}_x^2}{(\bar{p}_x)^2}\bar{B}_2^2-(\bar{B}_2)^2 
\label {C2Y} ; \quad
D_0=\frac{\mu_{ab}^2 a_x^2}{\bar{p}_x^2} \label {C2WW}
\end {eqnarray}
The factor $\frac{p_x}{\bar{p_x}}$ produces additional 
$\hat{\tau}_2$ fluctuations resulting from the particle velocity spread.
 without changing its expectation value.
Due to this effect absent in $C_q$ rotator model  
the part of $D(\tau_2)$ :
$$
D_x=D_0+\frac{\bar{p}_x^2-(\bar{p}_x)^2}{(\bar{p}_x)^2}(\bar{B}_2)^2\tau_0^2
$$
can be related to the packet smearing along $x$ coordinate, 
regarded as the clocks mechanism uncertainty. The realization of $x$
 measurement in ARF can be the intricated procedure, which
scheme we don't intend to discuss here. Some examples of the analogous
nonlocal observables measurements are described in \cite{Aha2}.

To calculate the time operator between two quantum RFs it's neccessary first
 to find the evolution equation for the  free motion in
quantum RF.  
 For the beginning we'll consider the evolution of system $S_2$ of
 RF $F^1$ and the neutral spinless particle $G^2$
which momentums $\vec{p}_i$ and energies $E_i$ are defined in 
 classical ARF.
If to regard  initially prepared states including only
positive energy components 
 , then their joint state vector evolution in 
ARF is defined by the sum of two  (KGR)
 Hamiltonians $\cite{Schw}$ :
\begin {equation}
-i\frac{d \Psi_s }{d\tau_0}
=[(m_1^2+\vec{p}_{1}^2)^{\frac{1}{2}}+(m_2^2+
\vec{p}_{2}^2)^{\frac{1}{2}}]\Psi_s \label {C0}
\end {equation}
 From it one should extract Hamiltonian
$\hat{H}^1$ of $S_2$ in $F^1$ rest frame which 
velocity relative to ARF is formally equal to 
$\vec{\beta}_1=\vec{p}_1E_1^{-1}$
Analogously to the calculations of the $G^2$ energy and  momentum
$\frac{s_{12}}{2} , \vec{q}_{12}$ in c.m.s.  by means of
Lorentz transformation with the parameter $\vec{\beta}_1$ written here
in vector form we define
 $S_2$ energy and $G^2$ momentum in $F^1$ rest frame  :
\begin {eqnarray}
  \vec{p}_{12}=\frac{s_{12}\vec{q}_{12}}{m_1}=\vec{p}_2+
\frac{(\vec{n}_1\vec{p}_2)(E_1-m_1)\vec{n}_1-E_2\vec{p}_1}{m_1}    \nonumber \\
  E^1_s=(s_{12}^2+\vec{p}^2_{12})^\frac{1}{2}=
 m_1+(m_2^2+\vec{p}^2_{12})^\frac{1}{2} \label {C2}
\end {eqnarray}
where $\vec{n}_1=\vec{p}_1 |\vec{p}_1|^{-1}$.  
 Yet $E_i , \vec{p}_i$ are the operators and their
transformations formally must be performed by the action of 
Poincare generators. Really it's easy to show that
  the transformation (\ref {C2}) can be 
 described as the generalization of Poincare group transformations
when its parameters $\vec{a},\vec{\beta}$ becomes the operators.
  The corresponding transformation operator is equal to :
\begin {equation}
       \hat{U}'_{A,1}=e^{i\vec{N}_2\vec{\beta}_1} 
  \label {C3A}
\end {equation}
where $\vec{N}_2=\frac{1}{2}(E_2\vec{r}_2+\vec{r}_2 E_2)$ is
Lorentz generator , $\vec{r}_2=i\frac{\partial}{\partial \vec{p}_2}$ ,
$\vec{\beta}_1$ is $F^1$ velocity operator defined above. Under this
transformation $\vec{p}_2 \rightarrow \vec{p}_{12}$ , 
$E_2 \rightarrow E_{12}=E^1_s-m_1$.
To obtain the evolution equation for $F^1$ proper time $\tau_1$
we'll assume that the operator relation
$-i\frac{\partial}{\partial \tau}=\hat{H}$ is applicable also for quantum RFs.
The resulting evolution equation for $G_2$  for $F^1$ proper time $\tau_1$ is :
\begin {equation} 
 -i\frac{d\psi^1}{d\tau_1}=
\hat{H}^1\psi^1=
[m_1+(m_2^2+\vec{p}^2_{12})^\frac{1}{2}]\psi^1
 \label {C4}
\end {equation}
It's easy to note that $\hat{H}^1$ depends only on relative motion of
$F^1,G^2$ and can be rewritten as function of $\vec{q}_{12}$.
 $\hat{H}^1$  coincides with KGR Hamiltonian, if  
  $m_1$ regarded as the arbitrary constant added to $G^2$ energy. 
 Consequently we can use in $F^1$ the same momentum eigenstates spectral 
decomposition and the states scalar product $\cite{Schw}$. This spectra
 can be used also as the basis of $G^2$ field secondary quantization
in $F^1$. To perform it we can introduce now antiparticles of $G^2$
, to which negative energy $E_{12}$ is attributed. Then taking the
square of eq. (\ref {C4}) , where $m_1$ is subtracted we can take
as the  field equation in $F^1$ coinciding with Klein - Gordon
equation :
\begin {equation}
  \frac{\partial^2\psi'^1}{\partial \tau_1^2}
=(m_1^2+\vec{p}_{12}^2) \psi'^1   \label {C4X}
\end {equation}

If the number of particles $N_g>1$ analogous to $\vec{p}_{21}$ of (\ref{C2}) 
canonical momentums can be defined for each particle separately.
 Alternatively for their states transformations from ARF to $F^1$  
the clasterization formalism can be used 
described here for  
 $N_g=2$  $\cite{Coe}$. According to previous arguments
 Hamiltonian in $F^1$ of two free particles $G^2,G^3$ rewritten through
the  system observables acquires the form   :
\begin {equation}
   \hat{H^1}=
   m_1+(s_{23}^2+\vec{p}^2_{23})^\frac{1}{2} \label {C7}
\end {equation}
,where $s_{23}$ is $G^2, G^3$ 
invariant mass.
 In clasterization
formalism at the first level the relative motion of $G^2, G^3$ 
defined by
$\vec{q}_{23}$ their relative momentum is considered.
 At the second level we regard them as the single quasiparticle
 - cluster $C_{23}$ with mass $s_{23}$ and $\vec{p}_{23}$  
  momentum in $F^1$ which evolution is studied. So at any level we regard 
 the relative motion of two objects only and this
 procedure can be extended in the obvious inductive way to  
  arbitrary $N$.

 As the space coordinate operator in $F^1$ 
  the generalization
of  Newton-Wigner ansatz $\cite{Wig}$ is natural to consider :
\begin {equation}
\hat{x}_{12}=i\frac{d}{dp_{12,x}}-i\frac{p_{12,x}}{2E_{12}^2} \label {C5}
\end {equation}

To obtain eq.($\ref{C4}$) only equivalence principle was used 
assuming that any quantum RF has its proper time without use of
any relation between $\tau_0$ and $\tau_1$ which will be studied now.
 In this framework for $F^2$ with $C_q$ clocks its
 Hamiltonian in $F^1$ can be found substituting
in $\hat{H}^1~$ $m_2=m'_{2}+\hat{H}_c$ and so we can find $\hat{\tau}_2$
solving Heisenberg equation for $\hat{H}^1$ analogously to 
 (\ref{C0H}). The similar
 calculations results in $F^2$ proper time operator
 $\hat{\tau}_2$ in $F^1$  :  
\begin {equation} 
    \hat{\tau}_2
= B_2 (\vec{p}_{12})\tau_1+\frac{\theta_{12}(0)}{2\pi\omega}  \label {C6}
\end {equation}
 Note that this approach is completely symmetrical and the
 operator obtained from(\ref{C6}) exchanging indexes 1 and 2
relates the
 time $\hat{\tau_1}$ in $F^1$ and $F^2$ proper time
- parameter $\tau_2$. Obtained relation between  two 
finite mass RFs shows that Quantum Equivalence principle
can be correct also in relativistic QM.
Analogously to Classical Relativity average time boost depends on whether
 $F^1$ measures $F^2$ clocks observables, as we considered or vice versa,
 and  this measurement  makes  $F^1$ and $F^2$ 
nonequivalent. 
 The  new effect will be found only
when $F^1$ and $F^2$ will compare their initially synchronized clocks.
Formally this synchronization means that at the moment $\tau^0_1$
the prepared $F^2$ state factorized as
 $\Phi^1_2(\vec{p}_{12}) \varphi_2(u_{in})$
 , where $\varphi_2$ is clock wave function , describing some
 initial time value
($|v_0\rangle$ for $C_q$ state).
If this experiment repeated several times
(to perform quantum ensemble) it'll reveal not only 
classical Lorentz  time boost ,
 but also the statistical spread having quantum origin with the
dispersion given in (\ref{C2X}).

Due to appearance of the  time operators
 the  transformation operator between two quantum RFs
$\hat{U}_{2,1}(\tau_2,\tau_1)$ is quite intricated, 
and to obtain it general form demands further studies,  here
 only the simplest situations are regarded.
 Consider first the transformation of  $F^2$ relative state in $F^1$ 
$\psi^1$ to $F^1$ state in $F^2$. The solution of eq. (\ref{C4})
is $\psi^1=\Phi_2(\vec{p}_{12})exp[-iE^1_s(\vec{p}_{12})\tau_1]$ and if
$F^1,F^2$ are sinchronized at $\tau_1=\tau_2=0$ then at this moment
one have $\Phi_1(\vec{p}_{21})=\hat{U}_{2,1}(0,0)\Phi_2(\vec{p}_{12})$.
It corresponds to RF Lorentz transformation $\vec{\beta}'_1=-\vec{\beta}_2$
which up to the quantum phase gives
 $\Phi_1(\vec{p}_{21})=\Phi_2(-\frac{m_1\vec{p}_{12}}{m_2})$.
Then it's easy to find that $\hat{U}_{2,1}(0,0)=\hat{C}_2\hat{P}_2$
the product of dilatation and parity operators as was shown obtaining
eq. (\ref{B3}). Then the transformation operator for any $\tau_1,\tau_2$ is :
 \begin {equation}
\hat{U}_{21}(\tau_1,\tau_2)=
\hat{W}_2(\tau_2)\hat{U}_{21}(0,0)\hat{W}_1^{-1}(\tau_1) \label{C8}
\end {equation}
, where  $\hat{W}_{1,2}(\tau_{1,2})=exp(-i\tau_{1,2}\hat{H}^{1,2})$
are the evolution operators in $F^1,F^2$. Analogously can be described the
transformation of the 
   single particle $G^3$ state between $F^1$ and $F^2$.
  To apply the clasterization formalism we'll take that
  in $F^1$  at time $\tau_1=0$ the joint state vector
   of $F^2$ and $m_3$ - 
is $\psi^1_{in}(\vec{p}_{23},\vec{q}_{23})=\sum c^1_{jk}|\vec{p}_{23,j}\rangle|
\vec{q}_{23,k}\rangle$ , where $\vec{p}_{23}$ is $F^2,G^3$ total
momentum in $F^1$.  
  Due to unambiguous correspondence  between the
   $\vec{p}_{13},\vec{q}_{13}$ and $\vec{p}_{23},\vec{q}_{23}$
phase space points
   the state vector $\psi^2_{in}(\vec{p}_{13},\vec{q}_{13})$ in $F^2$
is obtained acting on $\psi^1_{in}$ by 
 $\hat{U}_{2,1}(0,0)$.
Analytical relations connecting $\vec{p}_{13},\vec{q}_{23}$ and
$\vec{p}_{23},\vec{q}_{23}$ are quite complicated and omitted here
 $\cite{Byc}$.
 Then the joint $G^3,F^1$ state in $F^2$
at any $\tau_2$ can be obtained by the action of the operator
 $\hat{U}_{2,1}(\tau_1,\tau_2)$ of (\ref{C7})
on $G^3,F^2$ state in $F^1$.
 It means that despite $\tau_2$
and $\tau_1$ are correlated only statistically through $\hat{\tau}_2$
,  $G^3$ state vectors in 
$F^2, F^1$  at this moments are related unambiguously.

As was shown above  $G^2$ state  transformation
 (\ref{C2}) from ARF to $F^1$
can be described as the generalization of 
Poincare group transformations.
The  operator (\ref{C3A}) is equal to $\hat{U}_{A,1}(0,0)$   and the
calculation of $\hat{U}_{A,1}(\tau_0,\tau_1)$ becomes straightforward
in this case. This approach can be extended also on $F^2,G^3$ system
  analogously to transformation (\ref{C3A}). 
Combining our previous considerations
 we'll define in $F^1$ the transformation operator to $F^2$ :
\begin {equation}
  \hat{U}_{21}(0,0)=\hat{C}_2\hat{P}_2 e^{i\vec{\beta}'_2 \vec{N}'_3}
 \label{C9}
\end{equation}
where $\vec{N}'_3=\frac{1}{2}(E_{13}\vec{r}_{13}+\vec{r}_{13}E_{13}),\quad
 \vec{r}_{13}=i\frac{\partial}{\partial \vec{p}_{13}}, \quad 
 \vec{\beta}'_{2}=\vec{p}_{12} E_{12}^{-1}$.
Here $\vec{p}_{13} , E_{13}$ are defined for $G^3$ analogous to (\ref{C2}).
The first two members act on $F^2$ transforming it to
$F^1$ state , and the last part transforms $G^3$ state.

Now we'll consider obtained results in nonrelativistic limit.
It's easy to see that in the limit $\vec{p}_{12}\rightarrow 0$
 Hamiltonian (\ref{C2}) after the masses subtraction
differs from $\hat{H_c}$ of (\ref{B5}) by the factor 
$k_m=\frac{m_1+m_2}{m_1}$, resulting  from Lorentz transformation
 from c.m.s. to $F^1$ rest frame. The 
  space coordinate  operator in $F^1$ ${x}_{12}$ of (\ref{C5})
in nonrelativistic limit is equal to
$\hat{x}_{12}=k_m^{-1}(\hat{x}_2-\hat{x}_1)$ ,
 where $x_1,x_2$ are coordinates in ARF.
This result doesn't broke transformation invariance 
 , because nonrelativistic QM has no
fundamental length scale.

\section{Concluding Remarks}

  We've shown that the extrapolation of QM laws on free
macroscopic objects with which RF are associated
 demands to change the approach to the 
space-time  which was taken copiously from  Classical 
Physics. It seems that QM admits the existence of RF
 manifold each element of which is the state vector and
 the transformations between which principally can't
be reduced to Galilean or Lorentz transformations.

Historically QM formulation started from defining the wave functions on
Euclidean 3-space $R^3$ which constitute Hilbert space $H_s$.
 In the alternative approach
accepted here we can regard $H_s$ as primordial states
 manifold. Introducing
particular Hamiltonian  defines $\vec{r},~\vec{p}$ axes in $H_s$
and results in the asymmetry of $H_s$ vectors
 which permit
 to define $R^3$ as a spectrum of the continuous observable $\hat{\vec{r}}$
which eigenstates are
 $|\vec{r}_i>$. But as we've shown  here for several
 quantum objects one of which is RF
this definition
become ambiguous and have many alternative solutions defining $R^3$
on $H_s$. In the relativistic case the situation is more complicated, yet
as we've  shown it results in ambiguous Minkovsky space-time definition. 
Meanwhile each quantum $RF_i$ has its own proper time - parameter $\tau_i$
 and the phase space
and all this RFs are physically equivalent.  We have shown
that this parameteres can be related by the proper time operators ,
which introduces the quantum fluctuations in the time relations.
It means that to any time moment $\tau_i$  the time moment
$\tau_j$  in $RF_j$
can corresponds only with the uncertainty $\pm D^\frac{1}{2}(\tau_i)$. 
So in this model each observer has its proper space-time which can't be   
related unambiguously with the another observers space-time
 and in this sense is local. As the result we've got in any quantum RF
Hamiltonian Mechanics of free particles 
on fundumental 3-dimensional momentum space with time parameter.
 Due to the invariance of the obtained evolution
 equation (\ref{C4}) proposed Quantum Equivalence principle was
demonstrated  is applicable also in the relativistic case.

In our work we demanded strictly that each RF must be quantum observer
i.e. to be able to measure state vector parameters. But it isn't clear
 whether this ability is the main property
 characterizing RF. In classical Physics this ability 
doesn't influence the system principal dynamical properties. In QM at first
sight we can't
claim it true or false finally because we don't have the established theory
 of collapse. 
 But it can be seen from our analysis that collapse is needed
in any RF only to measure the wave functions parameters at some $t$.
Alternatively this parameters at any RF can be calculated given
the initial experimental conditions without performing the 
additional measurements.
It's quite reasonable to take that quantum states have objective meaning
and exist independently of 
their measurability by the particular observer,so this ability probably can't
 be decisive for this problem. It means that we can connect RF with the
system which doesn't include detectors ,which can weaken and simplify
our assumptions about RF. We can assume that primordial for  
 RF is the ability, which complex solid states have,
to reproduce and record the space and time points ordering  with
which objects wave functions are related. 

Author thanks M. Toller, V.Karmanov, V. Bykov for fruitful discussions. 

\end{sloppypar}

\end{document}